\def\dbcol{double column single spacing}
\ifdefined\dbcol
\ifdefined\ACM
	\documentclass{sig-alternate-10pt}
	\makeatletter
	\def\@copyrightspace{\relax} 
	\makeatother
\else
	\documentclass[conference,10pt,letter]{IEEEtran}
	\IEEEoverridecommandlockouts
	\makeatletter
	\def\ps@headings{%
	\def\@oddhead{\mbox{}\scriptsize\rightmark \hfil \thepage}%
	\def\@evenhead{\scriptsize\thepage \hfil \leftmark\mbox{}}%
	\def\@oddfoot{}%
	\def\@evenfoot{}}
	\makeatother
\fi

\usepackage[left=.6in,right=.6in,top=.5in,bottom=.5in]{geometry}
\else 
	\documentclass[conference,11pt,onecolumn]{IEEEtran}
\fi

\ifdefined\ACM 
	\usepackage{amsmath}
    \usepackage{cite} 
\else
	\usepackage[cmex10]{amsmath} 
	\usepackage{amsthm} 
	\ifdefined\COMSOC 
		\usepackage[nocompress]{cite}
	\else
		\usepackage{cite} 
	\fi
	\ifCLASSINFOpdf
	  \usepackage[pdftex]{graphicx} 
	\else
	  \usepackage[dvips]{graphicx}
	\fi
\fi

\ifdefined\META     \usepackage{stmaryrd}    \fi 

\newcommand{\begproof}{\ifdefined\dbcol\begin{IEEEproof}\else\begin{proof}\fi}
\newcommand{\Endproof}{\ifdefined\dbcol\end{IEEEproof}\else\end{proof}\fi}

\ifdefined\ACM
\else
	\usepackage[font=small]{caption} 
\fi
\ifdefined\COMSOC 
  \usepackage[caption=false,font=footnotesize,labelfont=sf,textfont=sf]{subfig}
\else
  \usepackage[caption=false,font=footnotesize]{subfig}
\fi

\usepackage{amssymb}
\usepackage{bm}
\usepackage{verbatim}
\usepackage{color}
\usepackage[normalem]{ulem}
\usepackage[ruled,lined,linesnumbered]{algorithm2e}

\newcommand{\fref}[1]{Fig.~\ref{#1}}
\newcommand{\tref}[1]{Table~\ref{#1}}
\newcommand{\sref}[1]{Section~\ref{#1}}

\newcommand{\cref}[1]{Corollary~\ref{#1}}

\newcommand{\aref}[1]{Algorithm~\ref{#1}}

\newcommand{\vect}[1]{\boldsymbol{\mathbf{#1}}}

\newcommand{\eps}{\epsilon}

\newcommand{\inv}[1]{\frac{1}{#1}} 

\usepackage{enumitem} 
\usepackage{bbm} 

\usepackage{fancyhdr}

\title{Manuscript ``Reshaping Mobile Crowd Sensing using Cross Validation to Improve Data Credibility''}
\author{
	\IEEEauthorblockN{Tie Luo\IEEEauthorrefmark{1} and Leonit Zeynalvand\IEEEauthorrefmark{2}\\
	\IEEEauthorblockA{
	\IEEEauthorrefmark{1}Institute for Infocomm Research, A*STAR, Singapore\\
	\IEEEauthorrefmark{2}School of Computer Science and Engineering, Nanyang Technological University, Singapore\\
	E-mail: luot@i2r.a-star.edu.sg, leonit001@e.ntu.edu.sg} } \vspace{-8mm} }

\begin{document}
\maketitle

\pagestyle{fancy}
\thispagestyle{fancy}
\chead{Accepted and to appear in: Proceedings of IEEE GLOBECOM 2017.}
\cfoot{}
\rhead{\small\thepage}
\renewcommand{\headrulewidth}{0pt}

\begin{abstract}
Data credibility is a crucial issue in mobile crowd sensing (MCS) and, more generally, people-centric Internet of Things (IoT). Prior work takes approaches such as incentive mechanism design and data mining to address this issue, while overlooking the power of crowds itself, which we exploit in this paper. In particular, we propose a cross validation approach which seeks a {\em validating crowd} to verify the data credibility of the original {\em sensing crowd}, and uses the verification result to reshape the original sensing dataset into a more credible posterior belief of the ground truth. 
Following this approach, we design a specific cross validation mechanism, which integrates four sampling techniques 
with a {\em privacy-aware competency-adaptive push} (PACAP) algorithm and is applicable to time-sensitive and quality-critical MCS applications. 
It does not require redesigning a new MCS system but rather functions as a lightweight ``plug-in'', 
making it easier for practical adoption. Our results demonstrate that the proposed mechanism substantially improves data credibility in terms of both reinforcing obscure truths and scavenging hidden truths.
\end{abstract}


\section{Introduction}\label{sec:intro}

The Internet of Things (IoT) ultimately boils down to the ``Internet of People'' in the sense that virtually all the services created by or via IoT have an ultimate goal of improving people's lives and experiences. In fact, IoT is not just a wired universe of sensors and devices; rather, it also connects people who in turn can play a critical role in the IoT ecosystem, by acting as data or service providers and not merely service consumers.

A key enabler of this vision is Mobile Crowd Sensing (MCS)\cite{guo15mcs_iot_si,mcs_iot16survey}, a people-centric sensing and computing paradigm that allows people who are equipped with mobile devices to act as sensors to collect a deluge of sensing data, over a large geographic area on a continuous time scale. 


However, despite several advantages (e.g., cost efficiency) of MCS as compared to ``traditional'' infrastructure-based IoT, one critical challenge faced by MCS is {\em data credibility} or {\em reliability}. That is, the quality of crowdsensed data can often be poor and inconsistent due to the unaccountability and the diverse nature of crowdworkers. Recent research efforts have attempted to address this issue from multiple angles, which broadly span from incentive mechanisms \cite{luo16tist,luo17commag,mobihoc15qoi_inc,sew14,kamar12inc_tr,science04bts,aaai13robust_bts} to assessment of quality and trustworthiness \cite{gc15crowd_trust,gc15endortrust,mswim10reputation}, and to truth discovery \cite{wang12truth,aamas15adaboost,shield15wisec}. However, existing solutions focus on analyzing the indigenous data or soliciting better quality data, while not fully utilizing the ``power of crowds'' \cite{power_crowd15science}.

In this paper, we address the data credibility problem by introducing a new {\em cross validation} approach into MCS. This approach does not mean comparing a piece of data (contributed by a worker\footnote{In this paper, we use ``worker'' and ``crowdworker'' interchangeably.} as in the MCS context) against another piece of data (contributed by another worker), which is similar to the approach used in the machine learning and data mining literature \cite{statlearn14book}. Instead, our proposed cross validation approach means subjecting the crowdsensed data to a group of crowdworkers who {\em did not} contribute to the original MCS data to seek their {\em verification}, for the purpose of {\em reshaping} or ``grooming'' the original sensed data toward the ground truth.

A counterargument, however, could be mounted as follows: rather than introducing such ``verifiers'', why not recruit another group of contributors to contribute more sensing data, which might also lead to a better representation of the ground truth? 
While this alternative sounds plausible, the rationale for our proposed cross validation approach are:
\begin{itemize}
\item MCS tasks typically have a {\em spatiotemporal constraint}, meaning that a worker must be at a specific geographic location during a specific time interval to collect a specified type of sensing data, which precludes many workers from being {\em eligible} contributors. However, there is a greater chance that a broader set of workers may be able to obtain relevant information from other channels (such as public media, his personal social network, domain knowledge or professional expertise). Therefore, these workers could evaluate the {\em quality} of data although they may be unable to contribute the actual {\em content}.
\item MCS tasks typically dictate certain effort from and incur sensing and communication costs to workers, 
and may entail revealing their private information (e.g., home or office locations) \cite{gc16privacy}. As a result, many workers may be reluctant to participate in MCS. However, it may be easier to convince them to simply provide a ``second-opinion'' about already obtained MCS data, and hence increase the likelihood of recruiting a large validating crowd in a short time. 
\end{itemize}

To summarize, our cross validation approach is justified by the observation that the contribution process entails restrictive and fairly demanding requirements whereas the validation process is much more lightweight and can amass much more evidence.

However, while the principle seems promising, designing a specific cross validation mechanism involves several key challenges: (i) Doesn't cross validation introduce another quality issue---the ``quality of validation''? In other words, how to ensure the {\em competency} of verifiers? (ii) How to expose the crowdsensed data to the validating crowd in an effective and efficient way? (iii) Since validation is to be performed by humans (rather than by comparing datasets \cite{statlearn14book}), how to minimize human {\em biases} and possible {\em privacy intrusion}? (iv) How to effectively {\em consolidate} the original crowdsensed dataset and the additional verification dataset? (v) How to provide incentives to verifiers? (How) should the consolidated outcome affect {\em contributors'} incentives? 

This paper addresses all the above challenges, and our contributions are summarized as follows.
\begin{enumerate}
\item We propose a new cross validation approach to address the data credibility problem for MCS. Unlike many prior works, it does not require redesigning MCS applications (which could jeopardize prior investment), but rather functions as a lightweight ``plug-in'' for {\em existing} MCS applications. For new MCS applications, the integration is even seamless.
\item We design a specific cross validation mechanism which integrates several sampling techniques with a progressive {\em privacy-aware competency-adaptive push} (PACAP) algorithm.  Corresponding to the challenges listed above, this mechanism is able to solicit {\em unbiased} opinions from verifiers in a {\em privacy-preserving} manner, and {\em reshape} the original MCS data by consolidating it with the verification dataset to form a more credible posterior belief model of the ground truth. Moreover, the mechanism also provides incentives for verifiers and adjusts incentives for contributors based on the reshaped outcome. 
\item The reshaped outcome (posterior belief) achieves significant improvement by being able to both reinforcing obscure (while identifiable) truths and scavenging hidden truths buried in the original MCS dataset.
\item Our mechanism is {\em practically viable} because: (i) it does not make strict assumptions on human rationality as most game-theoretical studies do, nor does it assume any specific (e.g., Gaussian) distribution of the underlying sensed phenomenon; (ii) it requires minimal effort from crowdworkers and is thus conducive to large-scale participation; (iii) it is simple to implement and operates fully autonomously. 
\end{enumerate}

\section{Related Work}\label{sec:relwork}

Data credibility is a crucial issue in the context of MCS and people-centric IoT. A large body of prior work has attempted to tackle this problem from different perspectives. 
One distinctive line of research is to influence crowdworkers' behaviors using {\em incentive mechanisms} such as auctions \cite{luo16tist} and trust and reputation systems \cite{luo17commag}, such that workers are compelled to contribute high-quality data. For example, Jin et al. \cite{mobihoc15qoi_inc} designed a quality-of-information aware incentive mechanism based on reverse combinatorial auction, to maximize social welfare which is a function of the quality of contributed data. A trust and reputation system called simple endorsement web (SEW), as proposed in \cite{sew14}, connects crowdworkers into a social network. 
SEW incentivizes high-quality contributions using inter-worker mutual benefits and a witness effect. Kamar and Horvitz\cite{kamar12inc_tr} designed an incentive mechanism with a consensus prediction rule to induce truthful reporting, but the mechanism only applies to crowdsourcing for a single correct answer. Bayesian truth serum \cite{science04bts,aaai13robust_bts} uses a scoring method that can remove bias in favor of consensus, but it requires each worker to explicitly predict the distribution of other workers' reports, which restricts its practicality. 

A different thread of research has focused on evaluating the {\em quality of crowdsourced data} or {\em trustworthiness of crowdworkers} so that more informed decisions can be made (e.g., on data or worker selection). Kantarci et al. \cite{gc15crowd_trust} treated crowdworkers to be part of a social network (based on the commonality of tasks assigned) 
to assess the trustworthiness of workers and data. 
Wu et al.\cite{gc15endortrust} proposed a reputation system that not only assesses but also predicts the trustworthiness of crowdworkers without requiring their prior contributions. 
Huang et al.\cite{mswim10reputation} used the Gompertz function to calculate device reputation score as a reflection of the trustworthiness of the contributed data. 

The third active area of research is {\em truth finding}, which aims to discover the ground truth by analyzing the noisy and possibly conflicting MCS data (sometimes taking workers into account as well), typically using data mining techniques. For example, Wang et al. \cite{wang12truth} uses the Expectation-Maximization (EM) algorithm to find the maximum likelihood estimate (MLE) of the probability that a given {\em binary} MCS measurement is true. Davami and Sukthankar \cite{aamas15adaboost} combined multiple trust-based data fusion techniques using AdaBoost, a machine learning algorithm, to predict whether a parking lot is occupied, based on crowdsourced user reports. 
Gisdakis et al. \cite{shield15wisec} proposed SHIELD to perform outlier detection in presence of compromised sensing devices, but it requires a large amount of data to train the complex machine learning model. 

Our proposed approach does not belong to any of these categories; instead, it further exploits the power of crowds \cite{power_crowd15science} by {\em introducing an extra (yet thin) layer of crowdsourcing on top of the original crowd-sensing}. 
Notably, it does not replace or exclude prior work such as above, but rather plays a complementary role by {\em reshaping} the original data so that existing methods (e.g., MLE) can still be applied (to the ``reshaped version'') to obtain better results.

{\bf Remark:} Our designed cross validation mechanism makes no assumption on whether the crowdsensed data must have binary (like \cite{wang12truth}), finite discrete, or continuous values, nor does it assume there must be a single (like \cite{kamar12inc_tr}) or multiple ground truth(s). In addition, unlike many statistical methods, it does not assume any specific (e.g., Gaussian) distribution of the underlying sensing phenomenon, nor does it assume any (common) prior held by crowdworkers. It also does not require workers to predict the distribution of other workers' reports as in \cite{science04bts,aaai13robust_bts}.


Our mechanism is also distinctly different from the peer voting or rating schemes used by online Q\&A forums (e.g., Quora and Stackoverflow) and product/service review websites (e.g., Amazon and TripAdvisor), in its suitability for time-sensitive and quality-critical MCS applications. We elaborate this in \sref{sec:sample}.

\section{Cross Validation Design}\label{sec:mechanism}

We present a systematic cross validation mechanism designed for MCS. It can be applied to general crowdsourcing platforms such as Amazon mTurk, CrowdFlower, and MicroWorkers, 
where many MCS campaigns are hosted. It can also be applied to a standalone MCS application where typically only a handful out of many registered workers are actually contributing in a specific spatiotemporal window. In both cases, there are a large pool of non-participants who we can exploit as the validating crowd.


For an overview of our cross validation mechanism, it first profiles the crowdsensed dataset in terms of its representative values and the probabilistic distribution of the data. Second, it presents the profile to the validating crowd using one of four sampling methods. 
Third, it strategically approaches the validating crowd using a PACAP algorithm to seek competent validation. 
Fourth, it reshapes the original MCS data profile using the validation feedback obtained above. Finally, it completes the loop by providing an incentive scheme for both the validating crowd and the original contributing crowd based on the reshaped outcome.

\subsection{Profiling}\label{sec:profile} 

A profile $\chi = (\mathcal V,\mathcal P)$ consists of a set $\mathcal V$ of {\em representative values} and a set $\mathcal P$ of {\em probability masses} corresponding to the representative values. The representative values are meant to represent the original crowdsensed data in a concise way while not losing practical precision.

To obtain this profile, the server that collating sensing data first creates a histogram of the crowdsensed data with an appropriate resolution (i.e., the bin width) determined by the specific MCS application. For example, in the case of road traffic speed, a bin width of 5mph would be a suitable choice; while for noise sensing, a bin width of 5dB would be reasonable. Clearly, this is not restricted by whether the data is continuously or discretely valued. 

Second, the server designates a representative value $v_i$ for each non-empty bin $i=1,2,...,n$, where $n$ is the total number of bins. For this we choose the {\em median} of each bin, while any other quantile of the data points, or the mean, of each bin can also be chosen provided that the resolution (determined in the previous step) is sufficiently high. Thus we obtain $\mathcal V=\{v_i | i=1,2,...,n\}$. 

Lastly, the server normalizes the histogram such that the {\em bin volume} $\kappa_i$---the number of data points contained in a bin---becomes a probability measure $p_i$. That is, $p_i=\kappa_i/\sum_{j=1}^n \kappa_j$. Thus we obtain $\mathcal P=\{p_i | i=1,2,...,n\}$.



Sometimes we also refer to $\mathcal P$ or $p_i$ as the {\em interim belief} so as to differentiate from the {\em prior belief} which is without observation of data; similarly we refer to the updated version of interim belief (after incorporating validation results in \sref{sec:reshape}) as the {\em posterior belief}. Now, we can say that $p_i$ is the probability that $v_i$ is the ground truth according to the interim belief. 

\subsection{Sampling}\label{sec:sample}

This procedure is to determine how to present the profile $\chi=(\mathcal V,\mathcal P)$ to the validating crowd. A candidate method is to expose $\chi$ or $\mathcal V$ at a public venue (e.g., a website) 
which is open to all workers, who may then pick the ``best'' representative values that they deem to be. This method is similar to what is used by many Q\&A forums such as Quora and Stackoverflow, e-commerce platforms such as Amazon, 
and recommendation sites such as TripAdvisor. However, it is too opportunistic by nature and not suitable for typically time-sensitive and quality-critical MCS applications, where we need to accumulate a sufficiently large number of validation inputs within a short timeframe. 

There are some alternatives, for example using subsets of values and/or workers, but they all have drawbacks (omitted due to space constraint) that eventually lead us to the following.

In our method, we present a {\em single} representative value to each chosen worker and ask him to give a single {\em rating} (e.g., ``Agree'' or ``Disagree'') to that value. This requires minimal effort from each worker, and allows for the flexibility of using fine-grained (i.e., multi-level) rating options. In this section, we address the problem of selecting representative values for effective validation, while the problem of selecting workers and approaching them for ratings is addressed in \sref{sec:select_rater}.

Our specific method of value selection is to {\em sample} each representative $v_i\in\mathcal V$ with a sampling probability $s_i$ such that each sampled value (not necessarily unique) will be presented to a (unique) rater. 
In particular, we consider the following sampling methods:
\begin{enumerate}
\item {\em Random Sampling}: $s_i=1/n$, so that all the representatives will be equally likely sampled. 
\item {\em Proportional Sampling}: $s_i=p_i$, so that a value that appears more often (hence more likely to be a truth) will be verified by more people.
\item {\em Reverse Sampling}: $s_i\propto d-p_i$ where $d$ is a constant. Hence by normalization we have
\begin{equation} 
s_i = \frac{d - p_i}{n d - 1}.
\end{equation}
Choosing a value for $d$ may appear straightforward since $1$, the upper bound on all probabilities, seems to be the natural choice. However, this will become problematic when all the $p_i$'s are fairly small (e.g., below 0.2), which is not uncommon. In that case, a value such as $d=1$ will {\em even out} the differences among all the $s_i$'s, and essentially equate reverse sampling to {\em random sampling}. Hence, we ``mirror'' $\mathcal P$ (as if $\mathcal P$ was plotted) with reference to its ``waistline'', $\frac{p_{min}+p_{max}}{2}$, where $p_{min}=\min_{p_i\in\mathcal P} p_i$ and $p_{max}=\max_{p_i\in\mathcal P} p_i$, and set $d=p_{min}+p_{max}$. 

\item {\em Inverse Sampling}: $s_i\propto 1/p_i$, so we have 
\begin{equation}
s_i = \frac{\inv{p_i}}{\sum_{i=1}^n \inv{p_i}}.
\end{equation}
\end{enumerate}

While random sampling and proportional sampling are most intuitive, we also explore reverse sampling and inverse sampling based on the following rationale:
\begin{itemize}
\item Values that frequently occur in the original crowdsensed dataset have already accumulated sufficient ``votes'' from crowdworkers and hence do not need as much validation as the less-occurring values. 
\item The real truth or the most-valuable information can sometimes be {\em uncommon} and thus buried in a haystack of common observations. However, conventional statistical methods generally filter out those minorities as {\em outliers} \cite{outlier10survey}. Reverse sampling and inverse sampling, on the other hand, usher in the opportunity of ``scavenging the hidden truths''.
\end{itemize}

The above sampling methods are compared in \sref{sec:eval}.

\subsection{Privacy-Aware Competency-Adaptive Push (PACAP)}\label{sec:select_rater}

In contrast to the passive approach taken by many e-commerce websites and Q\&A forums which ``wait'' for users to come to give feedback, we take a {\em proactive} approach to ``push'' rating tasks to a strategically chosen set of workers, whom we call {\em raters}, to seek their ratings. 

A push-based approach, however, faces two main issues. One is {\em privacy}, where improper (too frequent or irrelevant) pushes can be intrusive and annoying. The other is {\em competency}, where a chosen rater may not have the relevant information or domain knowledge to give a reliable rating, or may even purposely give false ratings.


In addition, there are also quantity requirements and time constraints we need to satisfy. Specifically, the {\bf problem statement} is to collect a desired number $m$ of {\em effective ratings}, under a {\em shortfall tolerance level} $\alpha$, within a timeframe $T_0$. Here, an effective rating is one that indicates a rater's positive or negative---instead of ``neutral''---preference (there is a good reason, though, to keep the ``neutral'' rating, which will be explained in \sref{sec:competency}). The shortfall tolerance level $\alpha\in(0,1)$ dictates that the any number below $m(1-\alpha)$ is unacceptable (while $m$ is more desirable). 
Moreover, the collected ratings are preferred to be {\em unbiased}. 

Our solution is a {\em privacy-aware competency-adaptive push} (PACAP) algorithm. We explain its general design principles in \sref{sec:bias}--3, and then present its specifics in \sref{sec:pacap}.

\subsubsection{\bf Anti-bias}\label{sec:bias}
Let us denote by $\mathcal U$ the set of all the workers registered on the MCS platform, and by $\mathcal C$ the set of contributors who have contributed to the original sensing task ($\mathcal C\subseteq \mathcal U$). We take three anti-bias measures. First, we exclude $\mathcal C$ from $\mathcal U$ to preclude contributors' biases toward their own respective contributions. Second, we ensure that each rater will receive at most one rating task. A rating task is composed of a representative value $v_i\in\mathcal V$, sampled according to \sref{sec:sample}, a task description, and a list of rating options.\footnote{For example, the task description could be ``Is the following value representative of the average traffic speed of Broadway, New York City during the morning peak hours today?'' or ``Is the following value representative of the average noise level at Lakewood Industrial Park between 8-10am today?'', and the rating options could be \{``Disagree'', ``Neutral'', ``Agree''\} or \{``Very unlikely'', ``Unlikely'', ``Not Sure'', ``Likely'', ``Very likely''\}.} Third, we mandate a {\em neutral} rating, which is explained below as it is also a measure of competency control.

\subsubsection{\bf Competency (quality) control}\label{sec:competency}
In order to collect competent ratings, we associate each worker $j=1,2,...,|\mathcal U|$ with a reputation $R_j$, which characterizes how accurate or reliable $j$'s past ratings were and is used as a proxy of $j$'s competency. The reputation $R_j$ is one of the (three) factors that determine each worker's opportunity of receiving {\em offers} (all the three factors as contained in \eqref{eq:prob_rater} will be explained in detail in \sref{sec:privacy}). An offer is a rating task; we call it this way because it can potentially lead to an increase in the recipient's reputation (detailed in \sref{sec:incentive}). But on the flip side, it can also lead to {\em penalty} if the rater's rating is {\em opposite} to the final verdict, in which case his reputation will be reduced. Therefore, we include a mandatory {\em neutral} rating option in every offer to allow a chosen rater to decline an offer without being penalized (e.g., if he deems himself not competent for that particular rating task). Therefore, by including both opportunity and risk as well as a neutral option, our offers prompt raters to only rate for tasks they are competent at, and to give unbiased ratings.

\subsubsection{\bf Privacy awareness}\label{sec:privacy}
Our push approach incorporates a privacy-aware competency-adaptive scheme, which selects a rater $j$ at time $t$ according to a probability $q_j(t)$ defined by
\begin{equation}\label{eq:prob_rater}
q_j(t) = \frac{1 - e^{- \lambda_j (t - t_j^-) (R_j+\eps)}}
{\sum_{j\in\Psi} \left[ 1 - e^{- \lambda_j (t - t_j^-) (R_j+\eps)} \right]}, \; j\in\Psi
\end{equation}
where $\Psi$ is a (dynamically updated) pool of workers from which we progressively search for raters, and
\begin{itemize}
\item $\lambda_j\ge 0$ is a {\em personalized elasticity parameter} catering for each worker's individual privacy preference,
\item $t_j^-$ is the last time when $j$ was given an offer, or the time when he signed up on the MCS platform if he has never been given an offer before,
\item $R_j\ge 0$ is $j$'s reputation described above, and
\item $\eps$ is a small positive constant (e.g., 0.1) to ensure that users with $R_j=0$ (such as new users) also have chance to receive offers.
\end{itemize}

The competency-adaptivity is explained by the fact that reputation $R_j$ in \eqref{eq:prob_rater} will determine each worker's probability of receiving offers. 
The privacy-awareness is accounted for by the following two factors. First, the exponent $(t - t_j^-)$ automatically spaces out the ``pushes'' to the same worker, which not only avoids too frequent pushes but also mitigates starvation by giving priority to workers who have not been offered a task for a long time. Note that, more reputable workers are still favored given the same waiting period. Second, the parameter $\lambda_j$ allows for {\em personalized privacy handling} as follows. It is initialized as a constant (say 1) for all the workers in $\mathcal U$. Then in each offer, we include three {\em optional} actions for the rater to choose: ``Send me more'', ``Send me less'', and ``Stop sending to me'', which correspond to updating $\lambda_j\gets\min(\lambda_j+\delta,\lambda_{max})$, $\lambda_j\gets\max(\lambda_j-\delta,\eps)$, and $\lambda_j\gets 0$, respectively. Here $\delta$ is a step size (e.g., 0.2), and $\lambda_{max}$ is a cap (e.g., 1.6) that prevents malicious users from abusing $\lambda_j$ to offset their low reputation $R_j$. Thus, we are able to accommodate individual privacy preferences while not jeopardizing the rater recruitment.

\begin{algorithm}[t]
\caption{Progressive PACAP}\label{alg:pacap}
\KwIn{\small Crowdworkers $\mathcal U$, contributors $\mathcal C$, representatives $\mathcal V$, target $m$, tolerance $\alpha$, deadline $T_0$}
\KwOut{$\mathcal R=\{\langle r_j(v_i), j, v_i\rangle | r_j(v_i)\neq 0, j\in\mathcal U, v_i\in\mathcal V\}$ with $|\mathcal R|\ge m\cdot (1-\alpha)$, or FAIL otherwise}
$\mathcal R\gets\emptyset, \Psi\gets\mathcal U\setminus\mathcal C$\\
$m(1)\gets m$, $M_Y(0)\gets 0$, $M_N(0)\gets 0$ \\
\For { $k\gets 1$ \KwTo $T_0/\tau$ } {
	select $m(k)$ raters, denoted by a set $\mathcal M(k)$, from $\Psi$ according to Eq.~\eqref{eq:prob_rater} \label{line:select_rater}\\
	\For {\rm{each} $j\in\mathcal M(k)$} {
		obtain one $v_i\in\mathcal V$ using a sampling method from \sref{sec:sample} \label{line:sample}\\
		wrap $v_i$ in a rating task and push it to rater $j$ to seek rating $r_j(v_i)$
	}
	wait for $\tau$ while collecting ratings:\label{line:collect1}
	$\circ\quad \mathcal R(k) \gets \{\langle r_j(v_i), j, v_i\rangle | r_j(v_i)\neq 0\}$\linebreak
	$\circ\quad m_N(k) \gets \sum_j \mathbbm 1_{r_j(v_i)=0} $\\
	$\mathcal R \gets \mathcal R \cup \mathcal R(k),  \quad m_Y(k) \gets |\mathcal R(k)|$\label{line:collect2}\\
	\If {$|\mathcal R|\ge m$ \label{line:target_meet}} {\Return $\mathcal R$ \tcp{SUCCESS *}}
	\tcp{Prepare for the next cycle:}
	update $\Psi\gets\Psi\setminus\mathcal M(k)$ \label{line:update_user}\\
	$M_Y(k)\gets M_Y(k-1)+m_Y(k)$, 
	$M_N(k)\gets M_N(k-1)+m_N(k)$,\label{line:responded}\\
	determine the scale of next outreach: \label{line:nextreach}
	\begin{eqnarray}
	m(k+1) \gets \left[m - M_Y(k)\right] \left[ 1+\frac{M_N(k)}{M_Y(k)} \right] \label{eq:nextreach}
	\end{eqnarray}
}
\uIf {$|\mathcal R|<m(1-\alpha)$ \label{line:finalcheck}} {\Return FAIL}
\Else {\Return $\mathcal R$ \tcp{SUCCESS}}
\end{algorithm}

\begin{table}[t]
\caption{Key notation for \aref{alg:pacap}}
\centering
\begin{tabular}{ c|l } 
\hline
 $m_Y(k)$ & no. of effective ratings collected in the $k$-th cycle \\  \hline
 $m_N(k)$ & no. of neutral ratings collected in the $k$-th cycle \\  \hline
 $m(k)$ & no. of offers sent out in the $k$-th cycle \\  \hline
 $M_Y(k)$ & no. of effective ratings collected up to the $k$-th cycle \\  \hline
 $M_N(k)$ & no. of neutral ratings collected up to the $k$-th cycle \\  \hline
\end{tabular}
\label{tab:notation} 
\end{table}

\subsubsection{\bf Progressive PACAP}\label{sec:pacap}
Given the above design principles, now we explain the entire algorithm.

In view of the time constraint $T_0$ and uncertain and dynamic worker behaviors, the algorithm divides $T_0$ into multiple cycles of smaller interval $\tau$ each, and performs {\em progressive push} over cycles. By progressive we mean that in each cycle the algorithm will approach a different group of workers of a different scale (group size) with a different number of offers, which are all re-calculated for each cycle by learning from previous cycles and adapting to the current. The rationale is to cater for the uncertainty and dynamics of worker responses (declination, acceptance, delay, or non-response).

The pseudo-code is given in \aref{alg:pacap} with key notation given in \tref{tab:notation}. In particular, we explain Line~\ref{line:nextreach} or Eq.~\eqref{eq:nextreach}, which describes how the algorithm determines $m(k+1)$, i.e., the number of raters to approach in the next (i.e., $(k+1)$-th) cycle. First, it notes that the number of effective ratings remaining to be collected is $m - M_Y(k)$. Second, it predicts the likelihood of obtaining an effective rating from a user in $\mathcal M(k+1)$ to be $M_Y(k) / \left[ M_Y(k) + M_N(k)\right]$, by assuming that the non-responding users of the current ($k$-th) cycle will respond (subsequently till $T_0$) with the same ratio as the users who have responded so far (till the $k$-th cycle). Thus, a division yields \eqref{eq:nextreach}.

\subsection{Reshaping}\label{sec:reshape}

Given two heterogeneous datasets, i.e., the original MCS data profile $\chi=(\mathcal V,\mathcal P)$ obtained in \sref{sec:profile}, and the set of effective ratings $\mathcal R$ obtained in \sref{sec:select_rater}, we consolidate them by reshaping $\chi$ using $\mathcal R$.

To this end, we first assign a score to each rating option contained in a rating task, as $\{-w_l,...,-w_2,-w_1, 0,\allowbreak w_1,w_2,...,w_l\}$ where $0<w_1<w_2<...<w_l$, for $(2l+1)$-level rating options. For example, 3-level rating options like \{``Disagree'', ``Neutral'', ``Agree''\} 
may be assigned scores $\{-1,0,1\}$, while 5-level rating options like \{``Very unlikely'', ``Unlikely'', ``Not Sure'', ``Likely'', ``Very likely''\} 
may be assigned scores $\{-2,-1,0,1,2\}$. 
In the following, without causing ambiguity, we slightly abuse notation by using $r_j(v_i)$ to denote a rating score as well as its corresponding semantic rating (e.g., ``Agree'').


Recall from \sref{sec:profile} that $p_i$ is the interim belief of how likely a representative $v_i$ is the ground truth, and that $\kappa_i$ is the bin volume before normalization. We reshape the original distribution $\chi$ by updating $p_i$ to the posterior $p'_i$ for all $i=1,2,...,n$, as follows. First, denote 
\begin{align}
g_i &=\inv{w_l}\sum_j r_j(v_i) \mathbbm{1}_{r_j(v_i)>0}, \label{eq:good}\\
b_i &= -\inv{w_l}\sum_j r_j(v_i) \mathbbm{1}_{r_j(v_i)<0} \label{eq:bad}
\end{align}
which are the number of normalized positive and negative ratings, respectively. For each $v_i$, the interim belief $p_i=\kappa_i/\sum_{j=1}^n \kappa_j$ can essentially be interpreted as $\kappa_i$ out of $\sum_{j=1}^n \kappa_j$ contributors have ``voted'' for $v_i$ to be the ground truth. Then during the cross validation phase, another $g_i$ out of $g_i+b_i$ raters voted for $v_i$. Note that it is not $g_i$ out of $\sum_{i=1}^n g_i + \sum_{i=1}^n b_i$ which is the total number of raters who have given effective ratings, i.e., $|\mathcal R|$. The reason is that the other $|\mathcal R| - g_i - b_i$ raters did not consider $v_i$ as a candidate at all since $v_i$ was not pushed to them. Thus, by considering both contributing and validating crowds, the belief $p_i$ could be reshaped as $\frac{\kappa_i + g_i}{\sum_j \kappa_j + (g_i+b_i)}$. 
However, this expression {\em biases} toward the larger group between the contributing crowd and the validating crowd $|\mathcal R|$. So we introduce an equalizing factor $\sum_{i=1}^n \kappa_i / |\mathcal R|$ to multiply $g_i$ and $g_i+b_i$, respectively, and thus the reshaped belief, before normalization, is given by
\begin{align}\label{eq:reshape_prob}
\hat{p_i} &= \frac{\kappa_i + \eta g_i \frac{\sum_{i=1}^n \kappa_i}{|\mathcal R|} }
	{\sum_{i=1}^n \kappa_i +\eta (g_i+b_i) \frac{\sum_{i=1}^n \kappa_i}{|\mathcal R|} }\nonumber\\
	&=\frac{p_i + \eta \frac{g_i}{|\mathcal R|} }{1+\eta \frac{g_i+b_i}{|\mathcal R|} }.
\end{align}
Here we have introduced a rescaling factor $\eta$ to allow weighting a (full-score) rating and an actual sensing contribution differently. For example, in a traffic monitoring MCS application, if the GPS sensors function as expected and the satellite signal is reliable, one could choose $\eta=0.5$ to down-weight the ratings; whereas if the accuracy of GPS sensors is in doubt (due to, e.g., urban canyons and tunnels), one could choose $\eta\ge 1$ to value the ratings more. 

The final posterior belief $p'_i$ is then obtained as
\begin{align}\label{eq:final_pi}
p'_i = \hat{p_i} \Big/ \sum_{i=1}^n \hat{p_i}.
\end{align}


We denote the reshaped profile by $\chi'=(\mathcal V, \mathcal P')$ where $\mathcal P'=\{p'_i|i=1,2,...,n\}$ is the set of posterior beliefs \eqref{eq:final_pi}.

\subsection{Incentive Scheme}\label{sec:incentive}

To motivate crowdworkers to participate in rating and contributing activities, we provide an incentive scheme that takes care of both aspects.


{\bf Raters:} For a rater $j$ who has given an effective rating $r_j(v_i)\neq 0$ for $v_i$, his reputation $R_j$ will be updated to $R'_j$ as
\begin{align}\label{eq:rep_update}
R'_j = [R_j + \Delta_j(v_i)]^+
\end{align}
where $[x]^+=\max(0,x)$, and
\begin{align}\label{eq:rep_delta}
\Delta_j(v_i) = \begin{cases}
	\frac{p'_i - p_i}{1 - p_i} \frac{r_j(v_i)}{w_l}, & \text{ if } p'_i > p_i \\
	\frac{p'_i - p_i}{p_i} \frac{r_j(v_i)}{w_l}, & \text{ if } p'_i < p_i. \end{cases}
\end{align}
The gist of the formulation \eqref{eq:rep_delta} is twofold:
\begin{enumerate}[label=G{\arabic*})]
\item Qualitatively, if a rating $r_j$ is {\em consistent} with the belief adjustment $p'_i - p_i$, i.e., a positive rating versus belief increment or negative versus decrement, then the rater $j$ will gain reputation. Otherwise, he will lose reputation which constitutes the {\em penalty} mentioned in \sref{sec:select_rater}. 
\item Quantitatively, the amplitude of reputation gain or loss is determined by two factors. One is the amount of belief adjustment $p'_i - p_i$ which measures the {\em impact} of cross validation on the original crowdsensing outcome. The other factor is the rating score $r_j$ which measures the extent to which the score has contributed to the above belief adjustment. Both factors are normalized, while the belief adjustment involves two cases ($p'_i \gtrless p_i$). 
\end{enumerate}

{\bf Contributors:}
A contributor $c\in\mathcal C$ who has contributed sensing data $v_i$ (or that which falls in the $i$-th bin) will receive payment as follows. Denote by $\pi_c(u_c, \vect u_{-c})$ the payment stipulated by the original incentive scheme\footnote{There is a rich literature on incentive mechanism design, for example \cite{luo16tist,mobihoc15qoi_inc,kamar12inc_tr,infocom15tullock,luo14mass}. For a comprehensive survey, see \cite{luo17commag,mcs_inc15jiot}.} of the MCS application, where $u_c$ is the quality of $c$'s contribution, and $\vect u_{-c}$ are the qualities of all the other contributors' contributions. 
Then after cross validation, the payment is revised from $\pi_c$ to $\pi'_c$ as
\begin{equation}\label{eq:payment}
\begin{split}\pi'_c &= \pi_c \left( u_c \frac{p'_i(c)}{p_i(c)}, \vect u'_{-c} \right), \\
\vect u'_{-c} &= \left\{u_{\tilde c} \frac{p'_{\tilde i}(\tilde c)}{p_{\tilde i}(\tilde c)} \bigg| \tilde c\in\mathcal C\setminus\{c\} \right\}
\end{split}
\end{equation}
where $p_i(c)$ and $p'_i(c)$ are an explicit version of $p_i$ and $p'_i$, respectively, that denote $c$ is the contributor. Hence \eqref{eq:payment} means that we keep the original incentive scheme $\pi$ intact while substituting all the qualities $u_c$ by $u_c \frac{p'_i(c)}{p_i(c)}$. The rationale is that, since $p_i(c)$ and $p'_i(c)$ are the likelihoods that $v_i$ is the ground truth under the interim and posterior beliefs, respectively, they can be treated as the quality and rectified quality of $v_i$, respectively.

Note that, however, the revised payment \eqref{eq:payment} may lead to the total payment not equal to the original total payment. Hence if one needs to satisfy a {\em fixed-budget constraint}, the following payment scheme can be used instead:
\begin{align}\label{eq:norm_payment}
\pi''_c = \frac{\pi'_c}{\sum_{\tilde c\in\mathcal C} \pi'_{\tilde c}} \sum_{\tilde c\in\mathcal C} \pi_{\tilde c}.
\end{align}

\section{Performance Evaluation}\label{sec:eval}

We consider a MCS-based IoT application in the transportation domain. It monitors the average traffic speed of a major road in a CBD area, by collecting GPS readings from participating crowdworkers who are drivers and passengers, via their smartphone- or car-borne GPS sensors. In practice, the quality of the sensor data are diverse and can be unreliable due to urban canyons, tunnels, errant driving behaviors, and faulty sensors. However, besides these workers who are the ``direct'' contributors, there are many other regular commuters who work in the CBD offices and occasional travelers who frequent the CBD from time to time for meetings. They may have experienced or witnessed the same traffic conditions as the MCS contributors do, and thus constitute a good candidate pool of validating crowd for our cross validation approach to tap on.

\subsection{Simulation Setup}

We have implemented our cross validation mechanism using Java. Our simulation generated a grand pool of $|\mathcal U|=50,000$ crowdworkers (e.g., those who have registered on a MCS platform such as Amazon mTurk). Among them, a subset of $|\mathcal C|=1000$ workers have contributed to the above CBD traffic monitoring MCS campaign during the morning peak hours of the day. Then in the cross validation phase, we aim to collect $m=1000$ effective ratings from the remaining 49,000 workers, with a shortfall tolerance $\alpha=0.1$, within $T_0=1$ hour.

To simulate worker behaviors, we use the following user model. A worker $j$ who is offered a rating task will accept the offer and give an effective rating with probability $a_j\sim U(0,1)$, and decline or ignore the offer or give a neutral rating with probability $1-a_j$. Each rater has an estimated truth $\nu_j$, which is a Gaussian random variable distributed as per $\mathcal N(\nu^*, 5)$ (unit: mph) 
where $\nu^*$ is the real truth. Upon accepting an offer, a rater $j$ compares the value $v_i$ specified in the offer with his estimated truth $\nu_j$, and 
rate 1 if $|v_i - \nu_j|\le TH_j$ and -1 otherwise (so $w_l=1, l=1$). Here $TH_j$ is $j$'s threshold which is also Gaussian and follows the distribution $\mathcal N(0.1\nu_j, 0.1\nu_j)$. 
Each rater responds to an offer with a delay that follows the exponential distribution of mean 15 minutes. In the reshaping step, $\eta=1$ (cf. Eq.~\ref{eq:reshape_prob}). 

\subsection{Results}

\begin{figure}[ht]
\fbox{\includegraphics[trim=0.6cm 0.2cm 2.6cm 3.7cm,clip,width=\linewidth]{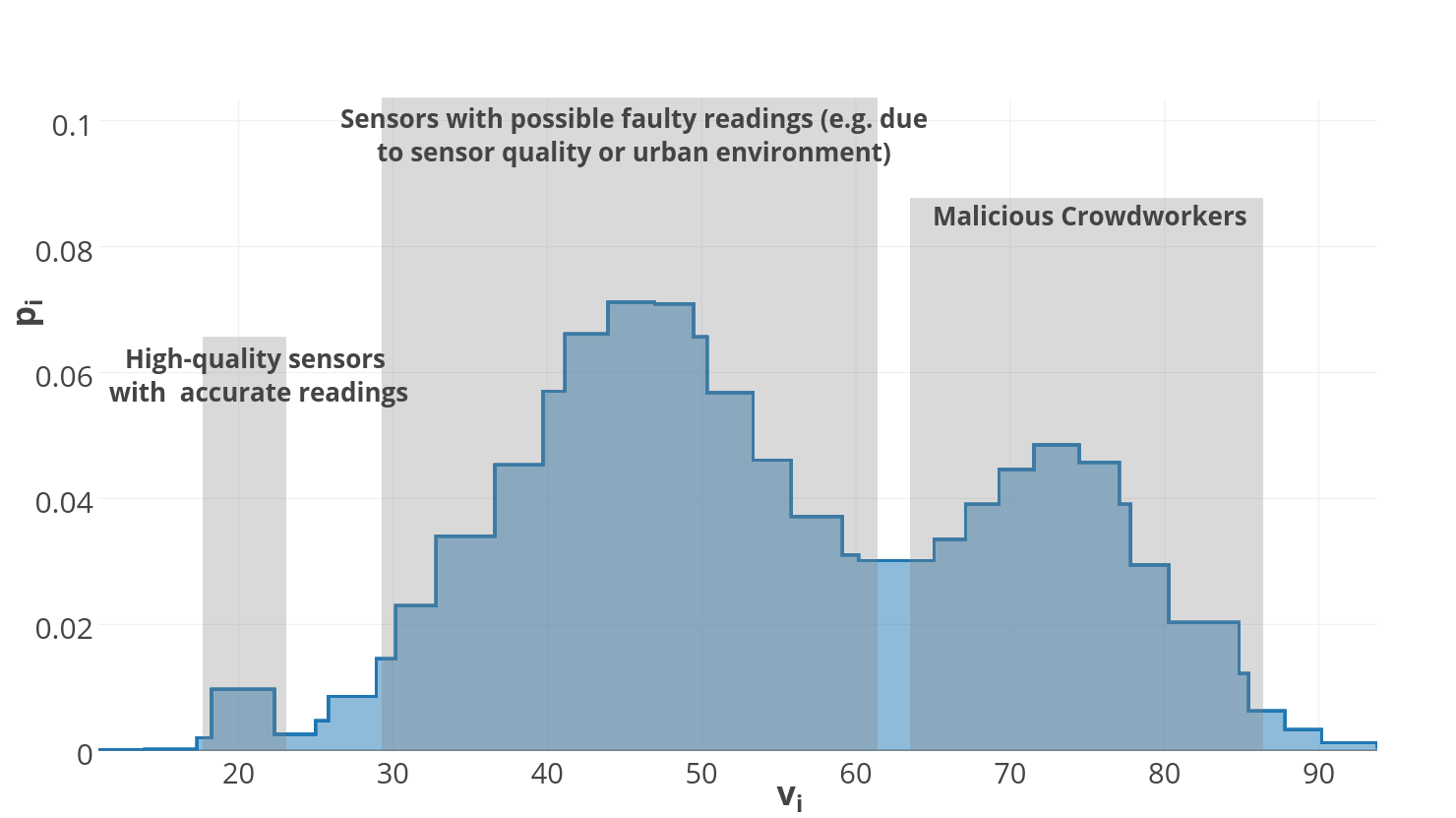}}
\caption{Profiling the original crowdsensed traffic speed data. The three annotations are for Case B only.}\label{fig:profile}
\end{figure}

The profile of the original crowdsensed data is presented in \fref{fig:profile}, which is obtained using the method described in \sref{sec:profile}. The crowdsensed traffic speeds range from 10 to 95 mph and the bin width is 5mph. While the profile seems to hint that the true traffic speed is likely 45 or 72 mph, the truth could also be 20mph because of the reason indicated in the three gray-shaded rectangular regions: the majority of the sensors yield inaccurate or faulty readings due to low sensor quality or urban environment, and only a handful provide high-quality readings. In the following, we refer to the former case which is ambiguous between 45 and 72 mph as Case A (obscure truth), and the latter case where the real truth 20mph is buried among noises as Case B (hidden truth).

\begin{figure}[ht]
\fbox{\includegraphics[trim=0.6cm 0.3cm 2.5cm 4.7cm,clip,width=\linewidth]{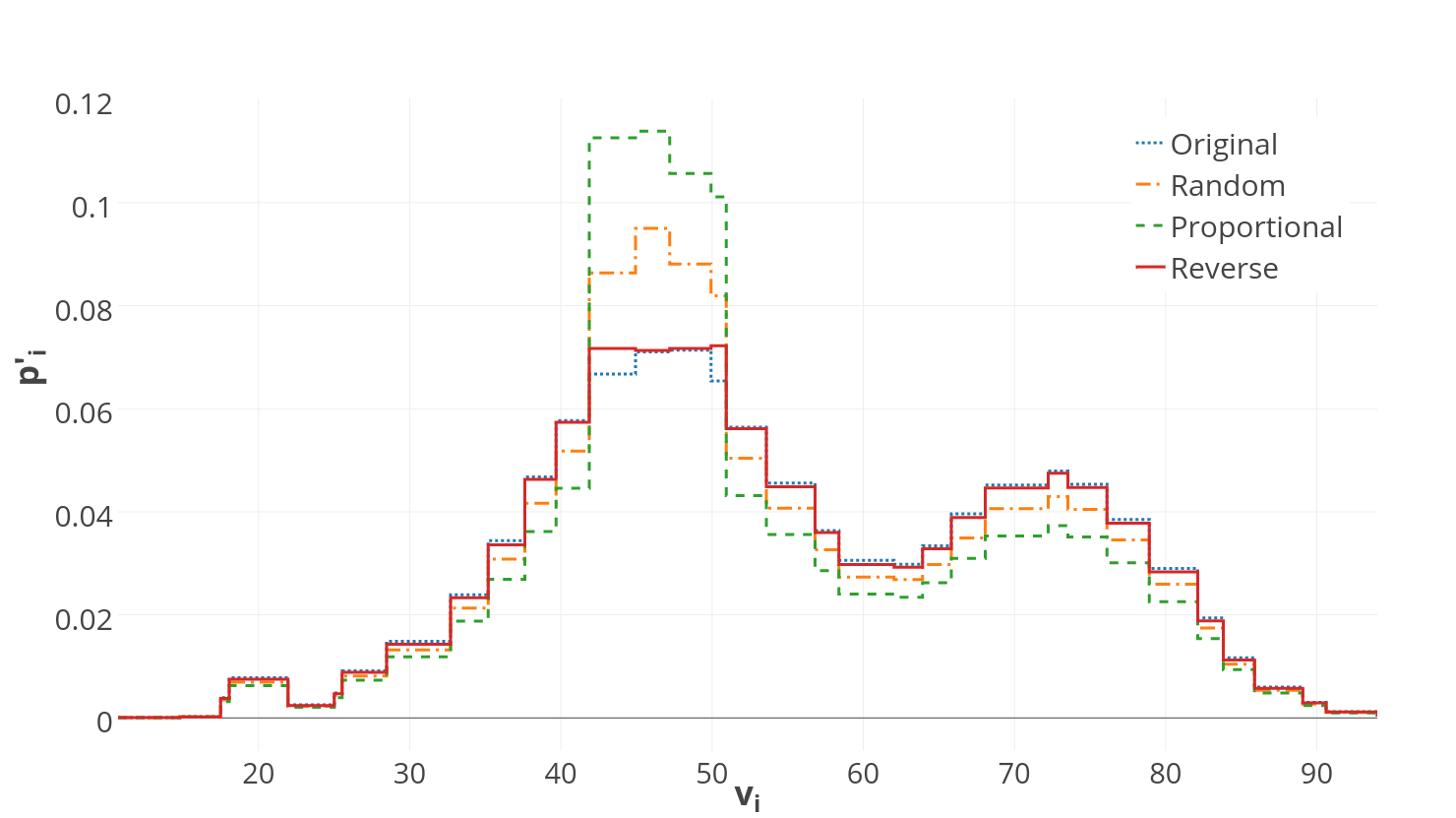}}
\caption{Truth reinforcement for obscure truths (Case A).}\label{fig:caseA}
\end{figure}

{\bf Truth reinforcement (Case A):}
In Case A, the consensus by majority vote does reveal the truth candidates but lacks of enough distinction. After applying our cross validation mechanism, the reshaped result is shown in \fref{fig:caseA}, where we see that all the sampling methods reinforce the real truth (45mph) by increasing its likelihood in the posterior belief. Proportional sampling performs the best in this case (a 75\% increment) because it leads to PACAP collecting the most ratings on values with the highest $p_i$, and hence amplifies the difference the most. Inverse sampling performs only slightly worse than Reverse sampling and is not plotted for visual clarity.

\begin{figure}[ht]
\fbox{\includegraphics[trim=0.5cm 0.3cm 2.5cm 4.7cm,clip,width=\linewidth]{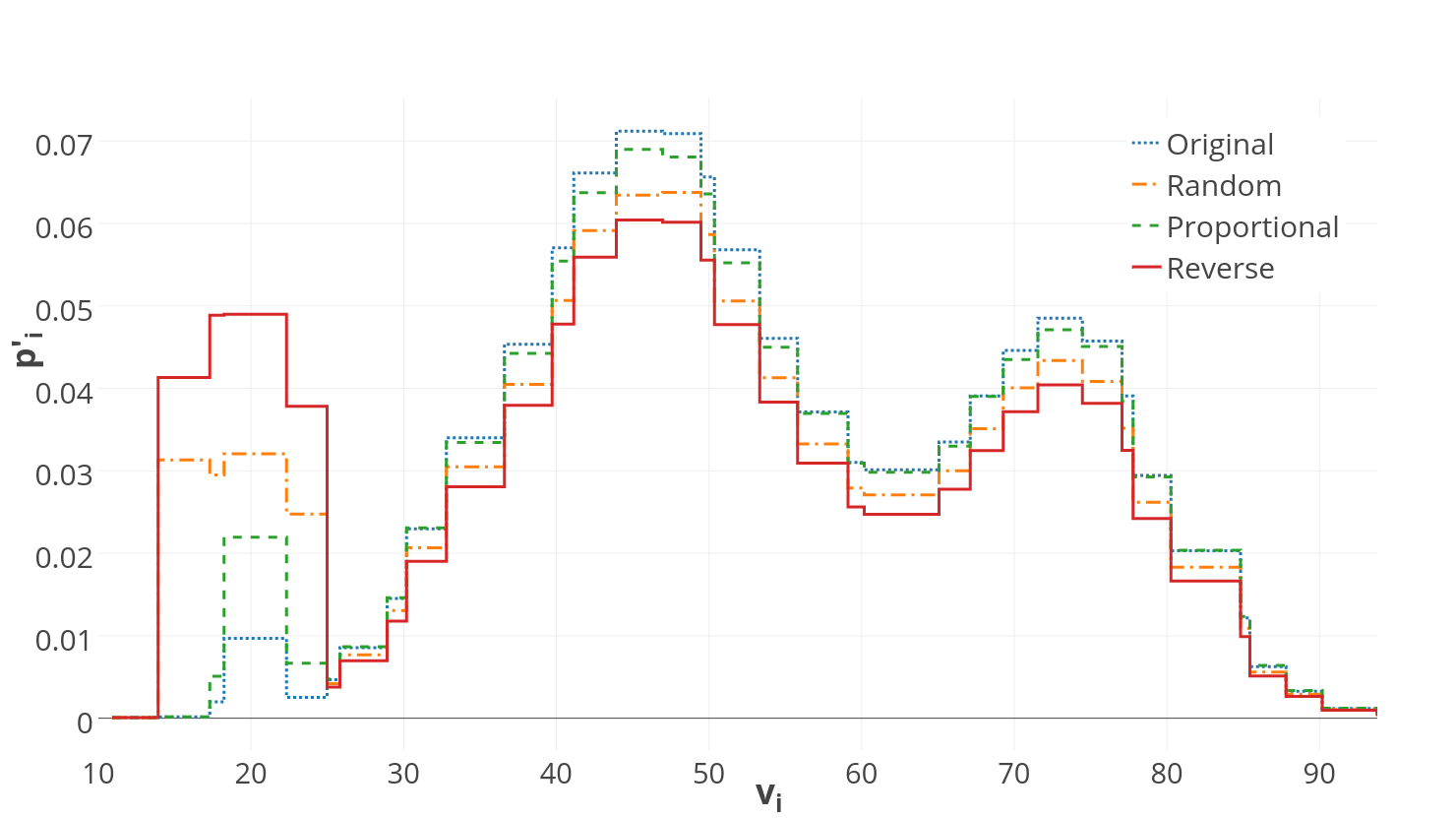}}
\caption{Scavenging the hidden truth (Case B).}\label{fig:caseB}
\end{figure}

{\bf Scavenging hidden truth (Case B):}
In Case B, the real truth is buried among inaccurate or false sensor readings, or due to unskilled workers. The reshaped outcome after applying our cross validation mechanism is shown in \fref{fig:caseB}, where Reverse sampling performs the best by boosting the likelihood of the real truth (20mph) to the maximum among all the three sampling methods (Inverse sampling performs marginally better). More specifically, the posterior $p'_i$ is a significant 4.9 times of the interim $p_i$ at $v_i=20$. In the meantime, the likelihood of the two false truths (45 and 72mph) are notably reduced. Consequently, the cross validation mechanism will either uncover the real truth unambiguously, or at least alert the administrator of the remarkable unreliability of the original crowdsensed data, by producing the drastic difference between the original and the reshaped profiles.

Overall, we would recommend Reverse sampling because it handles both cases A and B fairly well and strikes a reasonable trade-off between proportional sampling and inverse sampling.

{\bf Multiple truths and discrete data:}
The above results also show that our proposed mechanism is not restricted to single-truth scenarios but can apply to multi-truth scenarios as well. In addition, while the sensing data used in the simulation is continuous traffic speed, it can be seen (from profiling) that our mechanism works for discrete data as well. These two points are also mentioned in \sref{sec:relwork} (under ``Remark'').


\section{Conclusion}
To tackle the challenge of data credibility in MCS-based IoT applications, this paper proposes a new cross validation approach which further exploits the power of crowds by introducing an extra (yet thin) layer of crowdsourcing on top of the original crowd-sensing. Following the approach, we design a specific cross validation mechanism which integrates four sampling techniques with a progressive PACAP algorithm, and is well-suited for time-sensitive and quality-critical MCS applications. The mechanism does not make any assumptions on strict human rationality, the underlying distribution of sensed phenomena, or any prior held by workers, and can apply to both single-truth and multi-truth applications. It is also simple to implement and extremely lightweight (i.e., requires minimal effort from workers). 




\section*{Acknowledgment}
The authors would like to thank Andri R. Lauw who was involved in the earlier discussion of a related concept.

\bibliographystyle{IEEEtran}
\bibliography{IEEEabrv,CVCS}

\end{document}